\documentclass[11pt]{article}

\usepackage{amssymb, amsthm}
\usepackage{graphicx}
\usepackage{latexsym}
\usepackage{amsmath}
\usepackage{mathbbol}
\usepackage[utf8]{inputenc}
\usepackage{hyperref}

\begin{document}
\begin{titlepage}

\date{\today}
\begin{center}
{\bf\Large Phase space trajectories in quantum mechanics}

\vskip1.5cm

{\normalsize Christoph N\"olle}

\vskip1cm

\end{center}

\vskip1.5cm

\begin{abstract} 
An adapted representation of quantum mechanics sheds new light on the relationship between quantum states and classical states. In this approach the space of quantum states splits into a product of the state space of classical mechanics and a Hilbert space, and expectation values of observables decompose into their classical value plus a quantum correction, given a certain constraint on the initial conditions is satisfied. The splitting is then preserved under time evolution of the Schrödinger equation, and the time evolution of the classical part of a quantum state is governed by Hamilton's equation. The new representation is obtained from the usual Hilbert space representation of quantum mechanics by introducing a gauge degree of freedom in a time-dependent unitary transformation, followed by a non-conventional gauge fixing condition.	\end{abstract}
\end{titlepage}
\setcounter{tocdepth}{1}

 \tableofcontents

\section{Introduction}\label{sec:intro}
	A challenge in the interpretation of quantum mechanics consists in the difficulty to explain the emergence of classical physics in the so-called classical limit. The latter is often associated with a hypothetical limit $\hbar \rightarrow 0$ of Planck's constant $\hbar$. However, unlike for instance special relativity, where taking the limit $c\rightarrow \infty$ of the velocity of light in the equations of motion more or less straightfowardly leads to the Newtonian formulation of classical mechanics, the limit $\hbar\rightarrow 0$ is usually understood in a more symbolic way, since the mathematical model underlying quantum mechanics differs fundamentally from the model(s) of classical mechanics. In particular, the space of states for a point-particle in $\mathbb R^n$ looks very different in quantum mechanics ($\hbar \neq 0$) than in classical mechanics ($\hbar = 0$); pure quantum states are modeled by projective rays in an infinite-dimensional Hilbert space $\mathcal H = L^2(\mathbb R^n)$, whereas classical pure states are points in phase space $\mathbb R^{2n}$, interpreted as position and momentum vectors of the particle. 

\bigskip

	This apparent discrepancy has found a mathematical explanation in terms of $C^*$-algebras for observables, whose associated states are defined as normalized bounded linear functionals on the algebra. The latter is commutative in the classical case and acquires a non-commutative deformation in quantum mechanics. By means of the Gelfand representation, pure states on a commutative $C^*$-algebras can be identified with points in a classical phase space, and by means of the Gelfand–Naimark–Segal representation, pure states on a non-commutative $C^*$-algebra can be identified with (projective equivalence classes of) unit vectors in a Hilbert space.
	This relationship provides strong evidence that the two theories may indeed be connected by a limiting process. On the level of states the construction remains quite indirect, however.
	
	\bigskip
	
	Several approaches to the classical limit have been considered in the literature, among them the WKB approximation, the correspondence principle or large $N$ limit, the Wigner density, coherent states, stationary path integrals, deformation quantization and decoherence. While each of these offers some relevant insights into the classical limit, they all fall short of providing a simple and comprehensive derivation. See \cite{Werner95} for a concise review or \cite{Landsman05} for a more extensive review with historical and philosophical explanations. A recurrent theme is that the results apply to a limited set of special states (in particular for WKB, large $N$ and coherent states), often based on some ad-hoc constructions, or states are ignored completely, like in deformation quantization.

	\bigskip
	
	In this note it will be shown that the usual Schrödinger representation of quantum mechanics can be adapted by means of a time-dependent unitary transformation in such a way that it allows for a natural splitting of states and observables into a classical and a quantum part, and that the classical Hamiltonian equations of motions arise in this setting as a gauge fixing condition for the Schrödinger equation. The classical part of a quantum state, or observable, in this formalism is exactly the state, respectively observable, of the corresponding classical mechanics model in the Hamiltonian formalism, i.e. the time evolution of a pure state consists of a trajectory in phase space and the observable is modeled by a function on phase space.

	\bigskip

	The second part of the paper, starting with section \ref{sec:filtration}, deals with the classical limit. In a first step we will define a framework in which the limit $\hbar \rightarrow 0$ can be formulated. For this purpose we consider families of quantum states parametrized by $\hbar > 0$ and also promote $\hbar$ to an operator acting on the states by multiplication. Although this is apparently an unphysical model it allows us to formulate an exact condition on the initial state under which the limit $\hbar \rightarrow 0$ leads to the corresponding classical dynamics. It is shown that the splitting of states into classical and quantum contributions is then preserved under the time evolution of our adapted Schrödinger equation and that the time evolution of the classical part of a quantum states decouples completely from the purely quantum part. The proof is based on ideas from deformation quantization, namely Fedosov's filtration of the extended Weyl algebra. The intepretation of $\hbar$ as an operator is crucial here.
	
	\bigskip

	The results presented in the paper shed new light on the relation between the Schrödinger equation in quantum mechanics and Hamilton's equations in classical mechanics, with the latter appearing as a gauge fixing condition on the quantum state in our formalism. In addition, they enable us to formulate the conditions under which a quantum system allows for a semiclassical description and a classical limit. The interpretation of these latter results is not completely straight-forward, however, since they require us to consider families of states $\{\psi_\hbar\}$ for $\hbar > 0$, whereas our physical reality is bound to a fixed value $\hbar_0$. We will discuss some aspects of this in section  \ref{sec:interpretation}, without coming to a conclusive result. In particular, it remains open whether the "naive limit" approach $\hbar\rightarrow 0$ pursued in this note can in itself explain the emergence of classical physics from quantum mechanics. One possible approach to this problem could be to consider the large $N$-limit, for some quantum number $N$, where $\hbar$ can be identified with $1/N$. In a setting like this it may be possible to apply the results for a rigorous derivation of the classical limit. These applications are left for future work, however. 

	\bigskip

	Two sections on examples show our adapted formalism in action. One of them is dealing with the harmonic oscillator, the other with the hydrogen atom. It turns out that these two classical examples of quantum systems have quite different limiting behaviours. Whereas the eigenstates of the harmonic oscillator do satisfy the splitting condition, the hydrogen atom ground state does not. This implies that the latter cannot be described as a quantum perturbation of the classical system. The adapted quantum formalism is still applicable, but it does not single out any particular classical trajectory.

	\bigskip

	Appendix A gives a geometric interpretation for the transformed representation of quantum mechanics. It provides a connection to Fedosov's approach to deformation quantization, as well as to the geometric quantization framework. Based on this intepretation Appendix B discusses the generalization to curved phase spaces.

	\bigskip

	Throughout the paper we have to deal with unbounded, self-adjoint operators on a Hilbert space. In order not to distract from the core topic by technical details we will pretend that these operators were defined on the whole Hilbert space and mapped it into itself. On other occasions we will be imprecise about the exact function spaces to which certain results apply, as well, and furthermore ignore convergence questions of infinite series. Hence, the mathematical presentation is not rigorous. A popular approach to avoid the issues with self-adjoint operators is to only consider their bounded image under the exponential map. This strategy is not directly applicable here, however, since the quantum filtration on the algebra of observables introduced in section \ref{sec:filtration}, one of the core tools used, is not preserved under the exponential map.

\section{Textbook quantum mechanics}\label{sec:textbook}

	This paragraph serves to introduce our notation. We start with quantum mechanics on $\mathbb R^n$ in its textbook formulation, working in units where $\hbar$ is dimensionless ($\hbar \in \mathbb R_{>0}$). Let $q^1, \dots, q^n, p_1, \dots, p_n$ be coordinates on the classical phase space $\mathbb R^{2n}$. We use the symbol $y^\alpha$ ($\alpha=1,\dots, 2n$) to collectively refer to the $q^j,\ p_k$ coordinates, i.e. $y^j = q^j$ and $y^{n + j} = p_j$ for $j = 1, \dots, n$. Furthermore, in the context of the Hilbert space $\mathcal H := L^2(\mathbb R^n)$, we use a second set of coordinates $x^1, \dots, x^n$ on $\mathbb R^n$.
         On $\mathcal H $ we
	 have an action of the Weyl algebra $W$ generated by $\hat q^j$ and $\hat p_k$, defined as follows (where $j,k=1,\dots,n$ and $\psi \in \mathcal H$): 
	  \begin{equation} \label{canonicalOperators}
	    \hat q^j\psi(x)= x^j\psi(x), \qquad 
   	    \hat p_k\psi(x) = \frac \hbar i\frac \partial{\partial x^k}\psi(x).
	  \end{equation}
	 We will denote these generators collectively by $\hat y^\alpha$, ($\alpha=1,\dots,2n)$, i.e. $\hat y^j = \hat q^j$ and $\hat y^{n+j} = \hat p_j$ for $j=1,\dots, n$. Then the canonical commutation relations read
	 \begin{equation}\label{canonicalCommutationRelation}
	   [\hat y^\alpha,\hat y^\beta]=i\hbar\omega^{\alpha\beta},
	 \end{equation}
	 where $\omega$ is the symplectic form, concretely $(\omega^{\alpha\beta})_{\alpha,\beta = 1,\dots,2n} = \left(\begin{array}{cc}
		           0_{n\times n} & \mathbb 1_{n\times n} \\
		           -\mathbb 1_{n\times n} & 0_{n\times n}
		         \end{array} \right)$. 
To each smooth function $f(y)$ on phase space $M = \mathbb R^{2n}$ we associate an operator $\hat f$ on $\mathcal H$ by means of its Taylor expansion (summation over $\alpha_1, \dots, \alpha_k$ from 1 to $2n$ is understood):
	 \begin{align}\label{OperatorExp}
	  \hat f & = \sum_{k=0}^\infty \frac 1{k!} \big(\partial_{\alpha_1}
	  \dots \partial_{\alpha_k} f\big)(0) \hat y^{\alpha_1}\dots \hat y^{\alpha_k}  \\
	    &=f(0) +
	  (\partial_\alpha f)(0) \hat y^\alpha + \frac 12 \big(\partial_\alpha \partial_\beta f\big)
	  (0) \hat y^\alpha \hat y^\beta+\dots,\nonumber
	 \end{align}
	 Here $0$ denotes the origin in $\mathbb R^{2n}$. This is the quantum operator in Weyl-ordering, or symmetric ordering, for $f$. The expectation value of $\hat f $ in the state $\psi$ is given by the $L^2$-inner product $\langle \hat f \rangle_\psi := \langle \psi | \hat f \psi\rangle$, and the time evolution of a quantum state $\psi$ is governed by the Schrödinger equation
	 \begin{equation}\label{SchroedingerOriginal}
	 	i\hbar \partial_t \psi(t,x) = \hat H \psi(t,x),
	 \end{equation}
	 where $H \in C^\infty (M)$ is the Hamiltonian function of the system. 

	\bigskip

	Alternatively to the presented dynamics, it is possible to assign the dynamic behaviour of the system entirely to the observable and consider the state as time-independent. For this purpose, we define the time-dependent observable $\hat f_H(t) = e^{-\frac i\hbar t\hat H} \hat f e^{\frac i\hbar t\hat H}$, then the Schrödinger equation implies
	\begin{equation}\label{SchroedingerForObservable}
	 	i\hbar \partial_t { \hat {f \ }\!\!}_H(t) = [\hat H, { \hat {f \ }\!\!}_H(t)],
	 \end{equation}
	and the expectation value of the observable $f$ at time $t$ is
 	\begin{equation}
	 	\langle \hat f \rangle _{\psi(t)} = \langle \psi(t) | \hat f \psi(t) \rangle = \langle \psi(0) | \hat f_H(t) \psi(0) \rangle
	 \end{equation}

\section{The trajectory gauge}\label{sec:trajectoryGauge}
	For an arbitrary point $y = (q,p)$ in phase space we define a unitary operator on $\mathcal H$, the so-called Weyl operator (summation over $j=1,\dots,n$, respectively $\alpha, \beta=1,\dots, 2n$, is understood):
	  \begin{equation}\label{WeylOp}
	  U_y = U_{(q,p)} = \exp \Big[ \frac i\hbar \big(q^j\hat p_j - p_j\hat q^j \big)\Big] = \exp \Big[ -\frac i\hbar \omega_{\alpha\beta}y^\alpha \hat y^\beta \Big].
	 \end{equation}
	Explicitly, its action on a wave function in the position space representation \eqref{canonicalOperators} is
	\begin{equation}\label{WeylOperatorPosRep}
	  U_{(q,p)} \psi(x) = e ^{-\frac i\hbar p\cdot(x + \frac 12 q)}\psi(x+q).
	\end{equation}

	 For an observable $f\in C^\infty(M)$ we define the transformed operator $\tilde f_y$:
	 \begin{align}
	 	\tilde f_{y} &= U_{y} \hat f U_{y}^{-1} \nonumber \\
			&=  \sum_{k=0}^\infty \frac 1{k!} \big(\partial_{\alpha_1}
			  \dots \partial_{\alpha_k} f\big)(y) \hat y^{\alpha_1}\dots \hat y^{\alpha_k} \label{OperatorTrafoAlpha} \\
			&=f(y) +  (\partial_\alpha f)(y) \hat y^\alpha + \frac 12 \big(\partial_\alpha \partial_\beta f\big) (y) \hat y^\alpha \hat y^\beta+\dots,\nonumber
	 \end{align} 
	Symbolically, we can write this as $\tilde f_y = f(y + \hat y) $. In particular, for the coordinate functions $y^\alpha$ we have $\tilde y^\alpha = y^\alpha + \hat y^\alpha$.
	In the next step, we take this ansatz further and allow the unitary transformation to be time dependent. Let $I\subset \mathbb R$ be a (time) interval, and $c:I \rightarrow M$ be a differentiable trajectory in phase space. For $t\in I$ define the operator $U(t)$ as
	 \begin{equation}\label{unitaryDef}
	 	U(t) := U_{c(t)} = \exp \Big[ -\frac i\hbar \omega_{\alpha\beta}c^\alpha (t)\hat y^\beta \Big].
	 \end{equation}
	 as well as $\tilde \psi (t) := U(t) \psi(t)$ for $\psi \in C^\infty(I, \mathcal H)$, and $ \tilde f (t) := U(t) \hat f U(t)^{-1}$ for an observable $f \in C^\infty(M)$.
	 Due to the additional time dependence in $\tilde \psi$, the Schrödinger equation \eqref{SchroedingerOriginal} expressed in terms of $\tilde \psi(t)$ and $\tilde H(t) = U(t) \hat H U(t)^{-1}$ acquires an additional term. It reads:
	 \begin{align}\label{SchroedingerMod2}
	 	i\hbar \partial_t \tilde \psi (t,x) &= -i\hbar(\partial_t U(t))U(t)^{-1}\tilde \psi(t,x) + \tilde H(t) \tilde\psi(t,x) \\
			&= \omega_{\alpha\beta} (\partial_t c^\alpha(t)) (\hat y^\beta + \tfrac 12 c^\beta(t)) \tilde \psi(t,x) + \tilde H(t) \tilde\psi(t,x) , \nonumber
	 \end{align}
	where the result for $\partial_t U(t)$ can be obtained from the explicit form $U(t) = \exp \Big[ \frac i\hbar \big(q^j(t)\hat p_j - p_j(t)\hat q^j \big)\Big]$ by means of the Baker-Campbell-Hausdorff formula.
	Due to \eqref{OperatorTrafoAlpha} we can write \eqref{SchroedingerMod2} as 
	\begin{align}\label{SchroedingerMod3}
	 	i\hbar \partial_t \tilde \psi (t,x) &= \bigg[\omega_{\alpha\beta} (\partial_t c^\alpha) (\hat y^\beta + \tfrac 12 c^\beta)  + \sum_{k=0}^\infty \frac 1{k!} \big(\partial_{\alpha_1}  \dots \partial_{\alpha_k} H\big)(c(t)) \hat y^{\alpha_1}\dots \hat y^{\alpha_k}   \bigg] \tilde \psi(t,x) \\
		\label{SchroedingerMod4}
		&= \bigg[ H(c(t)) + \frac 12 \omega_{\alpha\beta} (\partial_t c^\alpha(t)) c^\beta(t) + \big((\partial_\alpha H)(c(t)) - \omega_{\alpha\beta} (\partial_t c^\beta(t)\big) \hat y^\alpha \\
		& \qquad\qquad\qquad\qquad +\sum_{k=2}^\infty \frac 1{k!} \big(\partial_{\alpha_1}  \dots \partial_{\alpha_k} H\big)(c(t)) \hat y^{\alpha_1}\dots \hat y^{\alpha_k}  \bigg] \tilde \psi(t, x) \nonumber
	 \end{align}
	This equation looks very similar to the original Schrödinger equation \eqref{SchroedingerOriginal}, except that the Hamiltonian function is evaluated in the point $c(t) \in \mathbb R^{2n}$ instead of $0$, and the zeroth and first order terms in the Taylor expansion of $H$ are modified.

\section{Gauge fixing}\label{sec:fix}
	Equation \eqref{SchroedingerMod4} is still fully equivalent to our original Schrödinger equation, for all choices of trajectory $c$. Hence, $c$ can be considered a gauge parameter. Equation \eqref{SchroedingerMod4} guides us at a particular choice for this trajectory, however. If we impose on $c$ the differential equation 
	\begin{equation}\label{CondHamilton1}
		\partial_t c^\alpha(t) = \omega^{\alpha\beta} \partial_\beta H(c(t)),
	\end{equation}
	then the first order term in $\hat y^\alpha$ in the new Schrödinger equation \eqref{SchroedingerMod4} vanishes, and the latter simplifies to 
	\begin{equation}\label{SchroedingerMod5}
		i\hbar \partial_t \tilde \psi (t,x) = \bigg[ H(c(t)) - \frac 12\big(\partial_\alpha H(c(t))) c^\alpha(t) + \sum_{k=2}^\infty \frac 1{k!} \big(\partial_{\alpha_1}  \dots \partial_{\alpha_k} H\big)(c(t)) \hat y^{\alpha_1}\dots \hat y^{\alpha_k}  \bigg] \tilde \psi(t, x)
	\end{equation}
	Note that condition \eqref{CondHamilton1} is nothing but Hamilton's equation of motion. If we write $c(t) = (q(t), p(t))$ and $\dot q = \partial_t q(t)$, etc., then it becomes
	\begin{equation}\label{CondHamilton2}
		\dot q^j = \frac {\partial H}{\partial p_j}, \quad \dot p_j = -\frac {\partial H}{\partial q^j}
	\end{equation}
	This appearance of Hamilton's equation(s) as a gauge fixing condition in quantum mechanics is quite astonishing, and we will see below how to interpret it. 
	
	\bigskip
	
	The expectation value of an observable $f$ becomes
	\begin{equation}\label{expectationValueClassicalExpansion1}
		\langle \hat f \rangle _{\psi(t)} = \langle \tilde f \rangle _{\tilde\psi(t)} = f(c(t)) + (\partial_\alpha f)(c(t)) \langle \hat y^\alpha \rangle_{\tilde\psi(t)}  + \dots,
	\end{equation}	
	where the first term on the right-hand side is exactly the classical expectation value of $f$ along the trajectory $c$. By imposing appropriate initial conditions on $c$ we should be able to choose it in such a way that it represents the corresponding classical state of our quantum system. In this case, all the higher order terms beyond $f(c(t))$ in \eqref{expectationValueClassicalExpansion1} should vanish in the classical limit. We will try to verify this observation in the remaining sections of the paper.

\section{The quantum filtration}\label{sec:filtration}

	In order to study the limiting behaviour for $\hbar \rightarrow 0$ of the Schrödinger equation \eqref{SchroedingerMod5} we will extend the algebra of observables and the state space in a way that allows us to treat $\hbar$ as a variable. Let us focus on the observables first. Consider the complex algebra $W_\hbar$ generated by purely formal symbols 
	\begin{equation}\label{filteredAlgebraGenerators}
		1,\ \hbar^{1/2},\ \hat y^\alpha,\ \hbar^{-1/2}\hat y^\alpha \quad (\alpha = 1,\dots, 2n)
	\end{equation}
	on which we impose that $1$ acts as identity, $\hbar^{1/2}$ commutes with everything, and the equivalence relations
	\begin{align} 
		\hbar^{1/2} \cdot \big(\hbar^{-1/2} \hat y^\alpha\big)  &\sim \hat y^\alpha  \\
		\hat y^\alpha \cdot \big(\hbar^{-1/2} \hat y^\beta\big)  &\sim \big(\hbar^{-1/2} \hat y^\alpha\big) \cdot \hat y^\beta \\
		[\hbar^{-1/2}\hat y^\alpha, \hbar^{-1/2}\hat y^\beta] &\sim i \omega^{\alpha\beta} 1
	\end{align}
	hold. Note that this also implies the canonical commutator $[\hat y^\alpha, \hat y^\beta] = i\hbar \omega^{\alpha\beta}$, if we write $\hbar$ for 
 	$(\hbar^{1/2})^2$. We can define the $\hbar$\textit{-degree} on this algebra by assigning degree $\frac 12$ to both $\hbar^{1/2}$ and $\hat y^\alpha$, and degree $0$ to $1$ and $\hbar^{-1/2}\hat y^\alpha$. Furthermore, let us introduce a filtration by defining the subspace $W_d \subset W_\hbar$ to be generated by all monomials in the generators \eqref{filteredAlgebraGenerators} of $\hbar$-degree at least $d/2$, for $d\in \mathbb N$. Note that the equivalence relations are compatible with the $\hbar$-degree and hence the filtration. We have $W_\hbar = W_0 \supset W_1 \supset W_2 \supset \dots$ , and the relation 
	\begin{equation}
		W_d \cdot W_e \subseteq W_{e+d}
	\end{equation}
	holds true for all $d,e\in \mathbb N$. 
	Now consider a time evolution of the form $i\hbar \partial_t \hat f(t) = [\hat A, \hat f(t)]$ on this algebra, for some fixed operator $\hat A \in W_\hbar$ and a time-dependent $\hat f: I \rightarrow W_\hbar$ (i.e. equation \eqref{SchroedingerForObservable}).
	We might be tempted to write this as 
	\begin{equation}\label{operatorEvolutionRearranged}
		\partial_t \hat f(t) = -\frac i\hbar \big[\hat A, \hat f(t)\big].
	\end{equation}
	However, the operator $-\frac i\hbar [\hat A, \cdot ]$ is not well-defined on $W_\hbar$ since the latter does not contain $\hbar^{-1}$. This would be incompatible with our filtration. In the special case that 
	\begin{equation}
		\hat A = \sum_{k=0}^\infty \frac 1{k!} A_{k; \alpha_1 \dots \alpha_k} \hat y^{\alpha_1} \dots \hat y^{\alpha_k}
	\end{equation}
	(with $A_{k; \alpha_1 \dots \alpha_k}$ totally symmetric in $\alpha_1, \dots, \alpha_k$, and possibly dependent on $\hbar^{1/2}$) does not contain a linear term:
	\begin{equation}\label{condLinearVanishing}
		A_{1;\alpha} = 0, \quad \text{for all } \alpha,
	\end{equation}
 	 we can nevertheless make sense of \eqref{operatorEvolutionRearranged}, since then 
	\begin{equation}
		-\frac i\hbar \big[\hat A, \cdot \big] := -\sum_{k=2}^\infty \frac i{k!} A_{k; \alpha_1 \dots \alpha_k} \big [(\hbar^{-1/2}\hat y^{\alpha_1})(\hbar^{-1/2}\hat y^{\alpha_2})\hat y^{\alpha_3} \dots \hat y^{\alpha_k},\ \cdot \ \big]
	\end{equation}
	Furthermore, this operator respects the filtration of the operator algebra. Note that for the original Hamiltonian operator $\hat H = \sum_k \frac 1{k!} \partial_{\alpha_1} \dots \partial_{\alpha_k} H(0) \hat y^{\alpha_1} \dots \hat y^{\alpha_k}$ of the Schrödinger equation \eqref{SchroedingerOriginal}, condition \eqref{condLinearVanishing} is violated (unless $H$ is constant), whereas for the modified Hamiltonian of equation \eqref{SchroedingerMod5} it is satisfied. This gives a first hint why the gauge condition \eqref{CondHamilton1} may be useful. 

\bigskip
	
	The quantum filtration on the operator algebra has been introduced in the context of deformation quantization, and has proved extremely valuable there \cite{Fedosov94}. We will denote $W_\hbar$ as the \textit{extended Weyl algebra}.

\section{The classical limit}\label{sec:filtrationApplic}
	
	In the previous section we have promoted $\hbar$ to an operator in the quantum algebra. Next, we would like to introduce an $\hbar$-dependency in the the state space as well. 
	So let's assume for now that we are given a smooth family $\psi_\hbar = \psi_\hbar(t_0)$ at fixed time $t_0$ of states in Hilbert space, parametrized by $\hbar > 0$. It is an element of $C^\infty(\mathbb R_{>0}) \otimes \mathcal H$. On this space we have an action of the operator algebra $W_\hbar$, which is generated by the canonical operators $\hat y^\alpha$ plus $\hbar$, the latter acting by multiplication. Let $H$ be a Hamiltonian function, $c$ a trajectory satisfying Hamilton's equation \eqref{CondHamilton1} with $c(t_0) = 0$, and $\tilde\psi(t) := U(t)\psi$ with $U(t)$ defined in terms of $c$ as in \eqref{unitaryDef} (we drop the $\hbar$ index on $\psi$ to avoid notational overload, but still consider $\psi$ to be paramterized by $\hbar$). These conditions imply that $\tilde \psi(t_0) = \psi$. We denote the operator from \eqref{SchroedingerMod5} by $\overset{\circ} H$:
	\begin{equation}\label{HamiltonianNew}
		\overset{\circ} H(t) :=  H(c(t)) - \frac 12(\partial_\alpha H(c(t))) c^\alpha(t) + \sum_{k=2}^\infty \frac 1{k!} \big(\partial_{\alpha_1}  \dots \partial_{\alpha_k} H\big)(c(t)) \hat y^{\alpha_1}\dots \hat y^{\alpha_k}.
	\end{equation}
	
	Then the time evolution of the expectation values of the canonical operators on $\tilde \psi(t)$ are
	\begin{align}
		\langle \hat y^{\alpha_1} \dots \hat y^{\alpha_k} \rangle_{\tilde\psi(t)} &=   \big\langle \psi(t_0) \big| e^{-\frac i\hbar \overset{\circ} Ht}\hat y^{\alpha_1} \dots \hat y^{\alpha_k}e^{\frac i\hbar \overset{\circ}{H}t} \psi(t_0) \big\rangle\\
			&= \big\langle \psi(t_0) \big| \big(e^{-\frac {it}\hbar [\overset{\circ}{H}, \cdot ]} \cdot (\hat y^{\alpha_1} \dots \hat y^{\alpha_k}) \big)  \psi (t_0)\big\rangle \nonumber
	\end{align}
	according to Baker-Campbell-Hausdorff. Now assume that the initial family $\psi(t_0) = \psi_\hbar(t_0)$ obeys the following regularity condition:
	\begin{equation}\label{filtrationCondInitial}
		\langle \psi(t_0) | \hat y^{\alpha_1} \dots \hat y^{\alpha_k} \psi(t_0) \rangle_\mathcal H = \mathcal O(\hbar^{k/2}) \quad \forall k \in \mathbb N,\ \alpha_1,\dots,\alpha_k =1, \dots ,2n,
	\end{equation}	
	where $\mathcal O(\hbar^{k/2})$ indicates the space of functions generated by monomials of degree at least $k$ in $\hbar^{1/2}$. This is equivalent to demanding the map 
	\begin{equation}
		W_\hbar \rightarrow C^\infty(\mathbb R_{>0}),\ \hat A \mapsto \langle \psi(t_0) | \hat A \psi(t_0) \rangle_\mathcal H
	\end{equation}
	to be filtration-preserving, where $W_\hbar$ carries its $\hbar$-filtration (see Section \ref{sec:filtration}) and the image of the map in $C^\infty(\mathbb R_{>0})$ is filtered by monomial degree (monomials in $\hbar^{1/2}$). We already know that the operator $-\frac i\hbar [\overset{\circ}{H}, \cdot ]$ preserves the filtration on the operator algebra, hence this is also true for its exponential, and we can conclude that \eqref{filtrationCondInitial} holds true for all times:
	\begin{equation}\label{hbarOrderPreserved}
		\langle \tilde\psi(t) | \hat y^{\alpha_1} \dots \hat y^{\alpha_k} \tilde\psi(t) \rangle = \mathcal O(\hbar^{k/2}) \quad \forall k \in \mathbb N,\ \alpha_1,\dots,\alpha_k =1, \dots ,2n.
	\end{equation}

	The expectation value of an observable $f$ in the state $\psi$, governed by the Hamiltonian $H$, is
	\begin{align}
		\langle \hat f \rangle_{\psi(t)} &= \langle  \tilde f (t) \rangle_{\tilde\psi(t)} \label{ExpValueEvolutionFinal}\\
			&= \underbrace{f(c(t))}_{\mathcal O(\hbar^0)} 
				+ \underbrace{\partial_\alpha f(c(t)) \big\langle \tilde\psi(t) | \hat y^\alpha \tilde \psi(t) \big\rangle}_{\mathcal O(\hbar^{1/2})} 
                                + \underbrace{\frac 12 \partial_\alpha \partial_\beta f(c(t))  \big\langle \tilde\psi(t) | \hat y^\alpha\hat y^\beta \tilde \psi(t) \big\rangle}_{\mathcal O(\hbar^1)}   + \dots\nonumber
	\end{align}
	Which implies that in the classical limit $\hbar \to 0$ the quantum mechanical expectation value $\langle \hat f \rangle_{\psi(t)}$ becomes equal to the classical expectation value $f(c(t))$:
	\begin{equation}\label{classicalLimitExp}
		\lim_{\hbar \to 0} \langle \hat f \rangle_{\psi(t)} = f(c(t)).
	\end{equation}

	The derivation of this result has been made possible by the representation in terms of $c$, $\tilde \psi, \tilde f$ and $\overset{\circ}{H}$, making use of the fact that $-\frac i\hbar [\overset{\circ}{H},\cdot]$ preserves the $\hbar$-filtration of the operator algebra. In principle, the textbook representation of quantum mechanics:

	\begin{equation}\label{ExpValueConstantC}
		\langle \hat f \rangle_{\psi(t)} = f(0)
				+ \partial_\alpha f(0) \big\langle \psi(t) | \hat y^\alpha \psi(t) \big\rangle 
                                + \frac 12 \partial_\alpha \partial_\beta f(0)  \big\langle \psi(t) | \hat y^\alpha\hat y^\beta \psi(t) \big\rangle   + \dots 
	\end{equation}
	must lead to the same result. However, in this case terms are not sorted by $\hbar$-degrees, so all of the infinite number of terms can contribute to the classical result at order $\mathcal O(\hbar^0)$, which makes it impossible to calculate the classical limit directly from \eqref{ExpValueConstantC}. It was the Hamilton equation \eqref{CondHamilton1} as gauge condition on $c$ that enabled us to sort the terms by $\hbar$-degree. Nevertheless, any other choice for $c$ is possible, and for a constant trajectory $c(t) = 0\ \forall t$ we get back the textbook formulation of quantum mechanics.

\bigskip

	Condition \eqref{filtrationCondInitial} is the relevant property that must be satisfied by the initial state of a quantum system in order for a semiclassical description to apply. Examples of families of wave functions that satisfy \eqref{filtrationCondInitial} are the eigenfunctions of the $n$-dimensonal harmonic oscillator, see section \ref{sec:harmosc} below. Particularly for these oscillator eigenfunctions and a specific class of Hamiltonians, the result \eqref{classicalLimitExp} has been obtained previously by Hepp \cite{Hepp74}, who also used the Weyl operators \eqref{WeylOp} in his derivation.

\section{Interpretation}\label{sec:interpretation}

	In the derivation of the limiting behaviour of observable expectation values \eqref{ExpValueEvolutionFinal} we had to make two assumptions:
	\begin{enumerate} 
		\item We are given a family of initial wave functions $\{\psi_\hbar(t_0)\}_{\hbar > 0}$.
		\item The family of initial wave functions satisfies \eqref{filtrationCondInitial}.
	\end{enumerate} 

	When modeling a physical system we normally demand the prescription of a single wave function $\psi_{\hbar_0} (t_0)$ as initial condition for the dynamical system. Here $\hbar_0$ denotes the physical value of the variable $\hbar$, and there is no physical principle that would allow us to extend the wave function to other values $\hbar \neq \hbar_0$. Hence, even the first assumption is not satisfied, and we cannot answer the question of how classical physics emerges in a quantum world purely within the formalism presented here. Instead, this will likely require an approach that takes into account the dimensionality of the constant $\hbar_0$ and possibly some consideration of a large $N$-limit, for some quantum number $N$. We leave this investigation to future work.
	
\bigskip

	On the other hand, it is still possible to consider the limiting behaviour for $\hbar \rightarrow 0$ for a family of theories parametrized by $\hbar>0$, if there is some natural way to define the family of initial wave functions $\{\psi_\hbar(t_0)\}_{\hbar>0}$. A good case for this is an eigenfunction $\psi_m$ of an observable operator $\hat f$:
	\begin{equation}\label{eigenstate}
		\hat f \psi_m = \alpha_m \psi_m
	\end{equation}
	for some $\alpha_m \in \mathbb R$. For a given physical observable $f \in C^\infty(\mathbb R^{2n})$ the operator $\hat f$ is an element of the Weyl algebra, but we can also interpret it as an element of the extended Weyl algebra $W_\hbar$, in which case equation \eqref{eigenstate} extends to a condition on a family of eigenfunctions $\psi_{m, \hbar}$ (one might call this the \textit{off-shell} formulation of \eqref{eigenstate} with regard to $\hbar$):
	\begin{equation}\label{eigenstate2}
		\hat f \psi_{m, \hbar} = \alpha_m(\hbar) \psi_{m, \hbar}.
	\end{equation}
	In order to validate the classical limit it then remains to verify condition \eqref{filtrationCondInitial} for the family $\{\psi_{m,\hbar}\}_{\hbar>0}$. We will pursue this approach in the investigation of the examples of sections \ref{sec:harmosc} and \ref{sec:hydrogen} below, where the states considered are eigenfunctions of the Hamiltonian. Note that \eqref{eigenstate2} is just a reinterpretation of the common procedure to solve the eigenvalue equation $\hat f \psi = \alpha \psi$ for generic values of $\hbar$ instead of a specific value $\hbar_0$.
	
\section{Example: $1D$ harmonic oscillator}\label{sec:harmosc}
	Let $q,p$ be coordinates on phase space $\mathbb R^2$ of a one-dimensional point particle. Consider the Hamiltonian
	  \begin{equation}
	    H(q,p) = \frac 12\big(p^2+q^2\big)
	  \end{equation}
	 (the harmonic oscillator Hamiltonian for mass $m=1$ and frequency $\omega=1$). Its corresponding quantum operator $\tilde H$, as defined in equation \eqref{OperatorTrafoAlpha}, is 
	 \begin{equation}
	   \tilde H_{(q,p)} = \frac 12\big(q^2+p^2\big) + q\hat q +p \hat p + \frac
	   12 \big(\hat q^2+\hat p^2\big).
	 \end{equation}
	 If we were to set $q=p=0$ here we would obtain the common representation as $\frac 12(\hat p^2 + \hat q^2)$. The modified operator $\overset{\circ} {H}$ from \eqref{HamiltonianNew} is
	 \begin{equation}
	   \overset{\circ}{H}_{(q,p)} =\frac 12 \big(\hat q^2+\hat p^2\big).
	 \end{equation}
	 Note how the modified zeroth order term $H(c(t)) - \frac 12(\partial_\alpha H(c(t))) c^\alpha(t)$ vanishes completely in this example, and we end up with the ordinary quantum harmonic oscillator Hamiltonian. In general, the correction to the zeroth order term cancels out the quadratic term in $H(c(t))$. 
	 Classical solutions $c: \mathbb R\rightarrow \mathbb R^2$ of Hamilton's equation
	  $$ \partial_t c (t)= \left(
		              \begin{array}{cc}
		                0 & 1 \\
		                -1 & 0 \\
		              \end{array}
		            \right) c(t)
	  $$
	 are of the form
	 \begin{equation}\label{OscTrajectory}
	  c(t) = \exp\Big\{ \left(
		              \begin{array}{cc}
		                0 & 1 \\
		                -1 & 0 \\
		              \end{array}
		            \right)t
	   \Big\} c(0) = \left(\begin{array}{cc}
		           \cos\ t & \sin\ t \\
		           -\sin\ t & \cos\ t
		         \end{array} \right)c(0).
	 \end{equation}
	 
	 The Schrödinger equation \eqref{SchroedingerMod5} becomes
	 \begin{equation}\label{HOSchroedinger}
	     i\hbar \partial_t \tilde \psi = \frac 12\big(\hat p^2+\hat q^2\big)\tilde \psi.
	 \end{equation}
	 As is well known, stationary solutions to this equation are labelled by $n=0,1,2, \dots$, so let $|n \rangle $ be the $n$-th oscillator eigenfunction, and $\tilde \psi (t) = |n(t)\rangle$ its time-dependent counterpart. Since we have $\langle n| \hat y^\alpha |n\rangle = 0$, according to \eqref{OperatorTrafoAlpha} the expectation value of the position observable is
	\begin{equation}
		\langle q\rangle(t) = \langle \tilde q(t) \rangle_{|n(t)>} = c^q(t) = \cos(t) c^q(0) + \sin(t)  c^p(0),
	  \end{equation}
	 and there are no quantum corrections at all to the center of mass motion. The energy is
	 \begin{equation}
	 	\langle H\rangle = \langle \tilde H \rangle_{|n(t)>} = \frac 12 \big( c^q(0) ^2 + c^p(0)^2\big) + \hbar \bigg(n+\dfrac 12\bigg),
	 \end{equation}

	 where the first term is the classical energy and the second is a quantum correction. In order to validate the limiting behaviour \eqref{filtrationCondInitial} let us introduce creation and annihilation operators:
	\begin{equation}\label{creationAnnihilationOps}
	\hat a = \sqrt{\frac 1{2\hbar}}\big(\hat q+i\hat p\big) ,\qquad
		  \hat a^\dagger = \sqrt{\frac 1{2\hbar}}\big(\hat q-i\hat p\big)
	\end{equation}
	They act on the eigenstates $|n\rangle$ as follows:
	\begin{align}
		\hat a^\dagger |n\rangle&= \sqrt{n+1}|n + 1\rangle\\
		\hat a |n\rangle&= \sqrt{n}|n - 1\rangle
	\end{align}
	This implies that condition \eqref{filtrationCondInitial} is satisfied for the states $|n\rangle$.

\bigskip

	For completeness, let us check what the solutions we just constructed look like in the textbook formalism of quantum mechanics. We have
	\begin{equation}
	  \psi(t)=U(t)^{-1}\tilde \psi(t) = \exp\Big[\frac i\hbar \Big(c^p(t)\hat q -c^q(t)\hat p\Big)\Big] \big|n(t)\big\rangle.
	 \end{equation}
	In terms of the creation and annihilation operators \eqref{creationAnnihilationOps} this reads
	 \begin{equation}
	  \psi(t) =  \exp\Big(ze^{-it}\hat a^\dagger - \overline z e^{it}\hat a\Big)|n(t)\rangle,
	 \end{equation}
	 with $z= c^q(0)+ ic^p(0) $. For $n = 0$ this type of wave function is called a \textit{coherent state}. Coherent states are known to resemble
	 classical solutions as closely as possible, which is compatible with the fact that in our adapted formalism their quantum contribution is the ground state.

	\bigskip

	The example of the harmonic oscillator is quite special, because the Hamiltonian contains only quadratic terms and hence the modified Schrödinger equation looks exactly like the ordinary Schrödinger equation. This is related to the fact that the quantum operator $\overset{\circ} H$ in this case preserves not only the $\hbar$-filtration on the operator algebra, but even the $\hbar$-degree itself. For more general, non-quadratic systems this will not be the case. Quadratic operators in the Weyl algebra are also special in that they span the so-called metaplectic Lie algebra (see also Appendix \hyperref[sec:appA]{A}).

\section{Example: hydrogen atom ground state}\label{sec:hydrogen}
	
	The hydrogen atom is governed by a Hamiltonian of the form 
	\begin{equation}\label{hydrogenHamiltonian}
	  H(\vec q,\vec p) = \frac {|\vec p|^2}{2\mu} - \frac \alpha {|\vec q|}
	\end{equation}
	for the motion of the electron, where $\vec q, \vec p \in \mathbb R^3$ are 3-dimensional vectors, $\mu$ is the reduced electron mass, and $\alpha > 0$ is a constant that determines the attractional electrostatic force between the positively charged nucleus and the negatively charged electron. In classical mechanics this setting is known as the Kepler problem. It is used to describe the motion of planets around a central star (neglecting interplanetary interactions), because the gravitational force has the same $1/r$ dependence on the distance as the electrostatic Coulumb force. Solutions to the classical Hamiltonian equations are restricted to a fixed 2-dimensional plane in $\vec q$-space, as a consequence of angular momentum conservation. They consist of closed curves in the form of ellipses (with energy $E<0$), including circles as a special case, as well as parabolas ($E=0$) and hyperbolas ($E>0$) as non-closed trajectories. The energy spectrum of the classical model is unbounded from below. It is well-known, however, that the classical model is insufficient to describe the hydrogen atom, since a rotating charge would be subject to rapid energy loss due to electromagnetic radiation, so that the electron would spiral into the nucleus very quickly.

\bigskip

	The quantum system corresponding to \eqref{hydrogenHamiltonian} is also a classical example treated in all quantum mechanics textbooks and courses, so it should suffice to briefly recapitulate its main properties. The energy spectrum of the system also contains both negative and positive values, corresponding to bound and unbound states, respectively, but now the spectrum is bounded from below and the negative part of the spectrum is discrete. The state of lowest energy is commonly denoted by $1s$. In the position space representation its time-independent part is given by

	\begin{equation}\label{hydrogen1s}
		\psi^{1s}(\vec x) = \frac 1{\sqrt{\pi a_0^3}} e^{-|\vec x|/a_0},
	\end{equation}
	where $a_0$ is the so-called Bohr radius: 
	\begin{equation}
		a_0 = \frac {\hbar ^2}{m_e \alpha}.
	\end{equation}
	
	An explicit calculation of the $L^2$ inner product reveals that
	\begin{equation}
		\langle \psi |\hat q_j^k \psi \rangle \sim \hbar^{2k} ,\qquad \langle \psi | \hat p_j^k \psi \rangle \sim \hbar^{-k}
	\end{equation}
	for $j=1,2,3$ and $k \geq 0$, implying that condition \eqref{filtrationCondInitial} is violated.
	This is actually not too surprising. The classical energy value of the phase space origin $(\vec q,\vec p) = (0,0)$ is minus infinity, whereas the quantum ground state is known to have finite energy. Hence, it is impossible to represent the quantum energy as a perturbative correction to the classical value. The singularity in the $1/r$-potential forbids any perturbative treatment, at least for states with a significant probability density near $r=0$.
	
	\bigskip

	Even if we do not get a perturbative expansion in powers of $\hbar$ in this case, it is still possible to represent the quantum system over a non-trivial classical trajectory $c(t) = (\vec q(t), \vec p(t))$. There is simply no preference for trajectories obeying the classical equations of motion in this case. 
	Since the original wave function \eqref{hydrogen1s} is radially symmetric and centered at the origin, we can immediately deduce from \eqref{WeylOperatorPosRep} that the the spatial expectation value $\langle \vec q \rangle$ of the wave function $\tilde\psi_{c(t)}$ will be centered at $-\vec q(t)$ for every timestamp $t$. An obvious choice for the trajectory is the circular motion around the nucleus at radius $a_0$, like in the Bohr(-Sommerfeld) model. The center of the wave function will then rotate around the nucleus as well, but with an 180° offset with respect to the classical motion:
	\begin{equation}
		|\tilde \psi^{1s}_{c(t)}(\vec x,t)| ^ 2 = \frac 1 {\pi a_0^3} e^{-2|\vec x + \vec q(t)|/a_0},
	\end{equation}
	where the classical particle position $\vec q(t)$ is
	\begin{equation}
		\vec q(t) = a_0
			\begin{pmatrix}
				\text{cos}(\omega t) \\
				\text{sin}(\omega t) \\
				0
			\end{pmatrix}, \quad \omega= \sqrt {\frac \alpha {\mu a_0^3}}.
	\end{equation}

\begin{figure}
\includegraphics[width=0.24\textwidth]{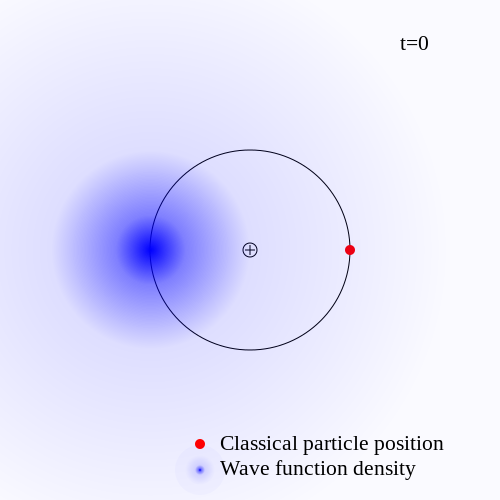}
\includegraphics[width=0.24\textwidth]{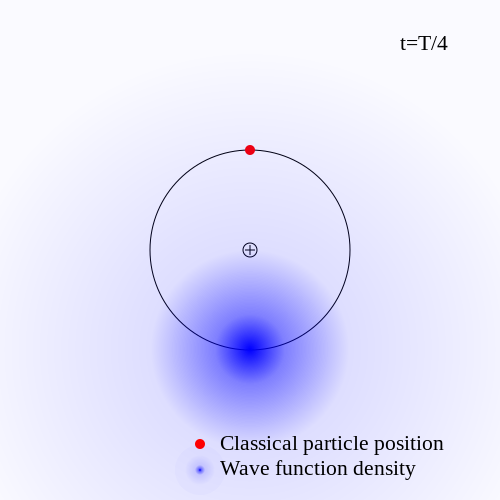}
\includegraphics[width=0.24\textwidth]{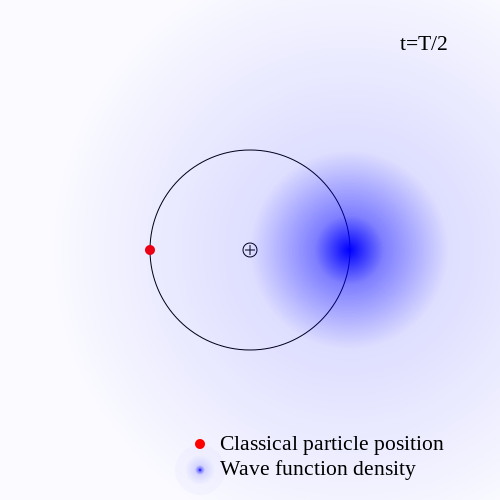}
\includegraphics[width=0.24\textwidth]{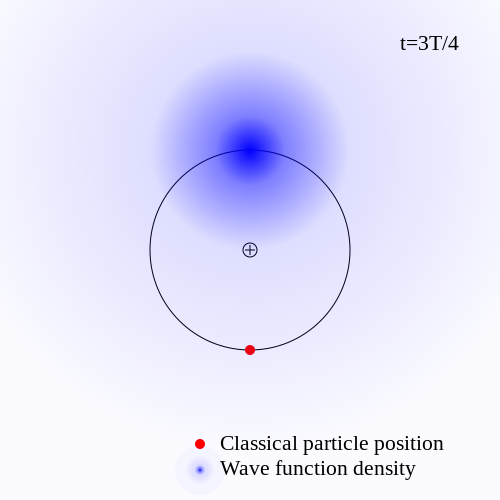}
\caption{The hydrogen ground state (1s) in a representation with a Bohr-Sommerfeld classical trajectory. The wave function is no longer rotationally symmetric, but its center rotates around the nucleus with a 180° offset with respect to the classical particle position. Note however, that due to the non-perturbative nature of the Coulomb potential no preference whatsoever can be deduced from our model for this particular classical trajectory. A similar representation is possible for every other trajectory in phase space, with the wave function always centered at minus the classical particle position.}
\label{fig:hydrogen-plots}
\end{figure}

	This is illustrated in Figure \ref{fig:hydrogen-plots}. Note that here the "classical" trajectory has a radius proportional to $\hbar^2$, so even if condition \eqref{filtrationCondInitial} was satisfied, our analysis of the splitting into classical and quantum contributions would not apply.

	\bigskip

	The hydrogen atom ground state nicely illustrates that condition \eqref{filtrationCondInitial} is not just a technicality, but it actually fails for systems that experience strong quantum effects. In this case classical solutions to the equations of motion do not seem to play any role in the quantum theory, and a classical limit does not exist.
	
\section{Summary}\label{sec:summary}
	We have defined a representation of quantum mechanics where every state consists of a pair ($c,\ \tilde \psi$), with $c: I \to \mathbb R^{2n}$ a trajctory in phase space and $\tilde \psi$ a wave function. An equivalence relation is defined on the set of all such pairs, which identifies pairs that can be transformed into each other by means of the transformations \eqref{unitaryDef}. The choice of trajectory $c$ to represent a certain quantum state can be thought of as a kind of gauge fixing, and the simplest choice $c(t) = 0\ \forall t$ gives us back the textbook representation of quantum mechanics. The general equation of motion is the modified Schrödinger equation \eqref{SchroedingerMod4} for $\tilde\psi$ and there is no restriction on $c$, but we found out that the Schrödinger equation simplifies if we select $c$ such that it satisfies Hamilton's equation \eqref{CondHamilton1}. In this case the quantum equation of motion is governed by the adapted Hamilton operator
	\begin{equation}
		\overset{\circ} H(t) =  H(c(t)) - \frac 12(\partial_\alpha H(c(t))) c^\alpha(t) + \sum_{k=2}^\infty \frac 1{k!} \big(\partial_{\alpha_1}  \dots \partial_{\alpha_k} H\big)(c(t)) \hat y^{\alpha_1}\dots \hat y^{\alpha_k}.
	\end{equation}
	 which is missing the linear term in the canonical operators $\hat y^\alpha$. This in turn implies that it preserves the quantum filtration on the operator algebra, which allowed us to deduce that the expectation value of an observable is given by the classical value of the observable plus quantum corrections, if the initial state satisfies our condition \eqref{filtrationCondInitial}. The latter ensures that quantum effects can be treated perturbatively, and it has been shown to be valid for the eigenstates of the harmonic oscillator but violated for the hydrogen ground state. 
	 
\bigskip

The table below summarizes the three equivalent representations of quantum mechanics defined above.

\begin{equation}
\begin{array}{ |c|c|c|c| }
	 \hline 
	       & \text{QM textbook} & \text{QM gauged} & \text{QM gauge fixed} \\ 
	 \hline
	 \text{State} & \psi &  [(c, \tilde \psi)] & (c, \tilde \psi) \\ 
	 \text{Observable} & \hat f & \tilde f  & \tilde f  \\
	 \text{Equation of motion} & i\hbar \partial_t \psi = \hat H \psi & \text{see \eqref{SchroedingerMod4}} & i\hbar \partial_t \tilde\psi = \overset{\circ} H(t)\tilde\psi \\ 
	 ~ &  ~ & ~ & \partial_t c^\alpha(t)=\omega^{\alpha\beta} \partial_\beta H (c(t)) \\
	 \text{Expectation value} & \langle \psi | \hat f \psi \rangle & \langle \tilde\psi | \tilde f \tilde\psi \rangle & \langle \tilde\psi | \tilde f \tilde\psi \rangle = f(c(t)) + \mathcal O(\hbar^{1/2})\\
	 \hline
\end{array}
\end{equation}

The relation between the quantities with and without tilde is given by $\tilde \psi(t) = U_{c(t)}\psi(t)$ and $\tilde f = U_{c(t)} \hat f U_{c(t)}^{-1}$, with the Weyl operator $U_{c(t)}$ defined in \eqref{unitaryDef}. The expression $[(c, \tilde\psi)]$ denotes the equivalence class of pairs $(c,\tilde\psi)$, where $c$ is a phase space trajectory and $\tilde\psi$ a wave function, and the equivalence relation is defined as $(c, \tilde\psi) \sim (d, U_d U_c^{-1} \tilde\psi)$. 

	\bigskip

	The explicit form of the gauge-fixed Schrödinger equation in the position space representation, for a typical Hamiltonian of the form
	\begin{equation}
		H(\vec q, \vec p) = \frac {|\vec p|^2} {2m} + V(\vec q),
	\end{equation}
	with some potential $V: \mathbb R^n \rightarrow \mathbb R$, reads
	\begin{equation}\label{SchroedingerNewExplicit}
		i\hbar \partial_t \tilde \psi(t, \vec x) = \bigg\{ -\frac {\hbar^2}{2m} \Delta + V(\vec q(t) + \vec x) - \vec\nabla V(\vec q(t) ) \cdot \vec x  \bigg\} \tilde \psi(t,\vec x),
	\end{equation}
	where $\Delta$ is the $n$-dimensional Laplace operator, $\vec \nabla V$ is the gradient of $V$, $\vec q(t)$ is the classical trajectory in configuration space (the particle position), and we have dropped some irrelevant zeroth order terms in the Schrödinger equation. We have seen in Section \ref{sec:harmosc} that the equation can be solved explicitly for the harmonic oscillator, but due to the explicit time dependence in the potential term, equation \eqref{SchroedingerNewExplicit} looks considerably more difficult to solve than the ordinary Schrödinger equation, in general. Hence, from a purely computational point of view the new representation may be rather useless, which could explain why it has not been considered previously. 

\section*{Appendix A: Geometric interpretation}\label{sec:appA}\addcontentsline{toc}{section}{Appendix A: Geometric interpretation}
	Consider a trivial vector bundle $\mathfrak H = \mathbb R^{2n} \times L^2(\mathbb R^n)$ over the phase space $M=\mathbb R^{2n}$, with fibre equal to the Hilbert space $L^2(\mathbb R^n)$. Sections of this vector bundle are functions $\psi \in \Gamma(\mathfrak H) = C^\infty(M) \otimes L^2(\mathbb R^n)$, i.e. functions $\psi_{(q,p)} (x)$, with $q,p,x \in \mathbb R^n$. As above, we use coordinates $y^\alpha$ with $\alpha = 1,\dots, 2n$ on the phase space that include both $q$ and $p$s. We can define a covariant derivative $D$ on the space of sections as follows:
 	\begin{equation}\label{FedCon}
	    D = d-\frac i\hbar \Big[\theta + \omega_{\alpha\beta}\hat y^\alpha dy^\beta\Big],
	  \end{equation}
	 where $d=dq^j\frac {\partial}{\partial q^j} + dp_j \frac {\partial}{\partial
	 p_j}$ is the exterior derivative, and $\theta$ is any 1-form on $\mathbb R^{2n}$
	 satisfying $d\theta=\omega = \frac 12 \omega_{\alpha\beta} dy^\alpha \wedge dy^\beta$. A convenient
	 choice is $\theta= \frac 12 \omega_{\alpha\beta}y^\alpha dy^\beta$, which we adopt. It can be shown that the covariant derivative is flat, i.e. its curvature form vanishes. This implies that we can find global solutions to the equation $D\phi = 0$.
         Explicitly, these solutions have the form
	 \begin{equation}\label{FlatSections}
	 	 \phi_{(q,p)}(x) = \chi(q+x) e^{-\frac i\hbar p(x+\frac q2)},
	 \end{equation}
	 where $\chi$ is any differentiable function. The operator corresponding to a function $f$, acting on the fibre $\mathcal
	 H_{(q,p)}$, is
	 \begin{align}\label{OperatorExp2}
		  \tilde f_{(q,p)}& = \sum_{k=0}^\infty \frac 1{k!} \partial_{\alpha_1}
		  \dots \partial_{\alpha_k} f(q,p) \hat y^{\alpha_1}\dots \hat y^{\alpha_k}  \\
		    &=f(q,p) +
		  \partial_\alpha f(q,p) \hat y^\alpha + \frac 12 \partial_\alpha \partial_\beta f
		  (q,p) \hat y^\alpha \hat y^\beta+\dots.\nonumber
	 \end{align}
	 Finally, the parallel transport operator $U:= U\big((q_0,p_0),(q,p)\big)$ for $D$, defined by $\phi_{(q,p)} = U \phi_{(q_0,p_0)}$ for the solutions \eqref{FlatSections}, is the Weyl operator
	 \begin{equation}
		  U = \exp \Big[ \frac i\hbar \Big((p_0-p)\hat q + (q-q_0)\hat p  +\textstyle{\frac 12} (qp_0-p q_0)\Big)\Big].
	 \end{equation}
	 Here summation over indices is understood, i.e. $pq$ means $p_j
	 q^j$, etc. The important property we need is that $U$ satisfies the
	 parallel transport equation
	  \begin{equation}
	    \partial_t U\big(y,c(t)\big) = -A_{c(t)}(\dot c(t))  U\big(y,c(t)\big),
         \end{equation}
	 where $y\in \mathbb R^{2n}$ is a point in phase space, $c$ is any curve starting in $y$, and $A$ is the connection form of $D$, i.e.
	  \begin{equation}
	    A = -\frac i\hbar \omega_{\alpha\beta} \Big[\frac 12 y^\alpha  + \hat y^\alpha \Big] dy^\beta.
	  \end{equation}
	 Consider the Schrödinger equation on the single fibre $\mathfrak H_y$. It reads $i\hbar \partial_t \psi(t) = \tilde H_y \psi(t)$. If we define 
	\begin{equation}
		\phi(t) := U\big(y,c(t)\big) \psi(t) \ \in\ \mathfrak H_{c(t)}
	\end{equation}		
	then the Schrödinger equation can be formulated for the flat section $\phi$:
	 \begin{align}
	  i\hbar \partial_t \phi &= i\hbar (\partial_t U) \psi + i\hbar U\partial_t  \psi \nonumber \\
		\label{SchroedingerOverCurve1}
	   &= -i\hbar A(\dot c)U\psi  + U\tilde H_y\psi \\
	   &= \big(\tilde H_{c(t)} - i\hbar A_{c(t)}(\dot c(t))\big)\phi(t),\nonumber
	 \end{align}
	 where we used that $\tilde H_{c(t)} = U\tilde H_y U^{-1}$. We could
	 now insert the explicit expressions for $\tilde H$ and $A$, but this
	 will not be very enlightening in general. Instead, we consider
	 special curves $c$, those who satisfy Hamilton's equation
	  \begin{equation}\label{HamiltonEq}
	   	\partial_t c^\alpha(t)  = \omega^{\alpha\beta} \partial_\beta H(c(t)).
	  \end{equation}
	 Then we get
	 \begin{equation}
		-i\hbar A(\dot c) = -\partial_\alpha H \hat y^\alpha - \tfrac 12 \partial_\alpha H y^\alpha,
	\end{equation}
	 and the first term on the rhs. cancels the terms linear in $\hat y^\alpha$ of $\tilde H$ in \eqref{SchroedingerOverCurve1}. Thus
	 \begin{equation}\label{SchroedingerCurveExplicit}
		  i\hbar \partial_t \phi = \Big(H-\frac 12 \partial_\alpha H y^\alpha + \sum_{k=2}^\infty \frac 1{k!}
		  \partial_{\alpha_1}\dots \partial_{\alpha_k} H \hat y^{\alpha_1}\dots
		  \hat y^{\alpha_k}\Big)\Big|_{c(t)} \phi,
	 \end{equation}
	This is again our equation \eqref{SchroedingerMod5}.

	\bigskip

	The phase space $\mathbb R^{2n}$ with its symplectic 2-form $\omega$ forms a symplectic manifold, and
	the vector bundle $\mathfrak H$ can be viewed as the bundle of symplectic spinors over $\mathbb R^{2n}$. There is an action of the symplectic group on $\mathbb R^{2n}$ 
	(the group of linear transformations preserving the symplectic form). The symplectic group has a universal covering group, the so-called metaplectic group. 
	The latter does not possess any finite-dimensional representation, but it can be represented on the Hilbert space $L^2(\mathbb R^n)$. Its Lie algebra is
	generated by symmetrized operators $\{\hat y^\alpha, \hat y^\beta\} = \frac 12 (\hat y^\alpha \hat y^\beta + \hat y^\beta \hat y^\alpha)$. Compare to the Spin group, whose Lie algebra is generated by the commutators of Dirac matrices [$\gamma^\alpha, \gamma^\beta$]. 

\section*{Appendix B: Generalization to curved phase space}\label{sec:appB}\addcontentsline{toc}{section}{Appendix B: Generalization to curved phase space}

	The construction presented in \hyperref[sec:appA]{Appendix A} generalizes to curved phase spaces. It is well-known that the Hamiltonian formulation of classical mechanics extends to symplectic manifolds $(M, \omega)$ of dimension $2n$, where $\omega$ is a non-degenerate, closed 2-form on $M$. In local coordinates $y^1, \dots, y^{2n}$ on $M$ we have $\omega = \frac 12 \omega_{\alpha\beta}(y)dy^\alpha \wedge dy^\beta$. And since $\omega$ is closed, $d\omega =0$, locally we can find a 1-form $\theta = \theta_\alpha dy^\alpha$ which satisfies $d\theta = \omega$. 
Let $c : I\subset \mathbb R \to M$ be a trajectory in phase space, then Hamilton's equation for $c$ reads:
	\begin{equation}
		\partial_t c^\alpha(t) = \omega^{\alpha\beta}(c(t)) \partial_\beta H(c(t)),
	\end{equation}
	where $H\in C^\infty(M)$ is the Hamilton function. The existence of the 2-form $\omega$ implies that there is an action of the symplectic group on the fibers of the tangent and cotangent bundles over $M$. The symplectic group $Sp(2n)$ is defined as the subgroup of all linear transformations on a symplectic vector space which leave the form $\omega$ invariant, i.e. transformations $U$ which satisfy $\omega_y (X,Y) = \omega_y(U\cdot X, U\cdot Y)$ for all $X,Y \in T_yM$. Now assume that the first Chern class $c_1(M)$ is even, and that the \textit{quantization condition}
	\begin{equation}
		\frac {[\omega]}{2\pi\hbar} \in H^2(M, \mathbb Z)
	\end{equation}
	for the cohomology class $[\omega]$ is satisfied. Then the $Sp(n)$-structure on the tangent space $TM$ can be lifted to an action of the metaplectic group on a Hilbert bundle $\mathfrak H$ on $M$, i.e. a vector bundle whose fibers are all isomorphic to the Hilbert space $L^2(\mathbb R^n)$. This is completely analogous to the lift of $SO(n)$-actions on the tangent bundle of an oriented Riemannian manifold to $Spin(n)$-actions on the associated spinor bundle, except that the spin representation is finite-dimensional and the metaplectic action is not. 

	It can be shown that on every symplectic manifold there is a torsion-free connection $\nabla$ on the tangent bundle (and associated bundles) which preserves the symplectic form $\omega$, i.e. $\nabla \omega = 0$. Contrary to the Riemannian case, this symplectic connection is not unique, but we simply choose an arbitray one. The connection lifts to the Hilbert bundle $\mathfrak H$, just like the spin connection lifts to the spinor bundle. In local coordinates we can write
	\begin{equation}
		\nabla_\alpha = \partial_\alpha + \Gamma_{\alpha\beta}^\gamma dy^\beta \otimes \partial_\gamma
	\end{equation}
	on the tangent space, and on $\mathfrak H$:
	\begin{equation}
		\nabla_\alpha = \partial_\alpha -\frac i{2\hbar}\Gamma_{\beta\gamma\alpha}\hat y^\beta \hat y^\gamma,
	\end{equation}
	where $\hat y^\alpha$ are the canonical operators acting on a fibre $\mathfrak H_y$, satisfying $[\hat y^\alpha,\hat y^\beta] = i\hbar \omega^{\alpha\beta}(y)$. Furthermore, $\Gamma_{\alpha\beta\gamma}:= \omega_{\alpha\delta} \Gamma_{\beta\gamma}^\delta$. In the previous section we chose our quantum states $\phi$ as parallel sections of the vector bundle $\mathfrak H$ over $\mathbb R^{2n}$. The most obvious choice would be to impose $\nabla \phi = 0$, for sections $\phi \in\Gamma(\mathfrak H)$. However, the connection $\nabla$ in general has a non-vanishing curvature, which implies that the equation $\nabla \phi =0$ does not possess any solutions. Therefore, we first need to tweak the connection a little. Fedosov has shown that by adding higher order terms in the operators $\hat y^\alpha$ it is possible to make the curvature form vanish projectively \cite{Fedosov94}. Fedosov's connection $D$ assumes the form
	 \begin{equation}\label{FedosovConnection}
		D_\alpha^{\mathfrak H} = \partial_\alpha + \frac i\hbar\omega_{\alpha\beta}\hat y^\beta -\frac i{2\hbar}\Gamma_{\beta\gamma\alpha}\hat y^\beta \hat y^\gamma 
				 - \frac i{8\hbar}R_{\beta\gamma\delta\alpha} \hat y^{\{\beta} \hat y^\gamma \hat y^{\delta\}}  + \mathcal O(\hbar ^{1})
	\end{equation}
	where $R_{\alpha\beta\gamma\delta} = \omega_{\alpha\kappa}R_{\beta\gamma\delta}^\kappa$ are the components of the curvature form of $\nabla$, and $\hat y^{\{\alpha} \hat y^\beta \hat y^{\gamma\}}$
	denotes the totally symmetrized product of the three operators. Higher order terms in the connection are determined by a recursive formula, which can be found in \cite{Fedosov94}. In the derivation of this 
	result Fedosov makes heavy use of the quantum filtration introduced in section \ref{sec:filtration}, which treats operators $\hat y^\alpha$ as having quantum level $1/2$, like $\hbar^{1/2}$. It should be noted that in general nothing can be said about the convergence of the series \eqref{FedosovConnection}, which is why Fedosov is very careful to define it only on some operator space of formal series in $\hbar^{1/2}$ and the $\hat y^\alpha$, similar to the one defined in section \ref{sec:filtration}. We'll pretend instead that \eqref{FedosovConnection} was well-defined, just to see where this leads us, but should be aware that the remainder of this section is mathematically ill-founded for generic symplectic manifolds.
	
	\bigskip
		
	The curvature form $F_{\mathfrak H} \in \Gamma (\Lambda^2 T^*M \otimes L(\mathfrak H)) $ of $D^{\mathfrak H}$ actually does not vanish completely, but is equal to 
	\begin{equation}\label{FedosovCurvature}
		F_{\mathfrak H} = \frac i\hbar \omega \otimes \mathbb 1_\mathfrak H, 
	\end{equation}
	where $\mathbb 1_\mathfrak H$ denotes the fibre-wise identity operator on $\mathfrak H$. So we are not quite there yet. Enter \textit{geometric quantization}: similarly to deformation quantization, geometric quantization was born out of an attempt to explicitly construct the quantum theory associated to the Hamiltonian mechanics on $(M, \omega)$, but with a focus on the states instead of the observables. An important ingredient in this construction is the so-called \textit{pre-quantum bundle} $B$. This is a line-bundle, i.e. a complex 1-dimensional vector bundle on $M$, which carries a connection $\nabla_B$. In a local trivialization of $B$ this pre-quantum connection can be written as 
	\begin{equation}
		\nabla_B = d - \frac i\hbar\theta,
	\end{equation}
	and its curvature form $F_B$ is 
	\begin{equation}\label{PrequantumCurvature}
		F_B = -\frac i\hbar \omega \otimes \mathbb 1_B. 
	\end{equation}
	Equations \eqref{FedosovCurvature} and \eqref{PrequantumCurvature} let us deduce that the product bundle $\mathfrak H \otimes B$ carries a flat connection 
	\begin{equation}
		D := D^{\mathfrak H} \otimes \mathbb 1_B + \mathbb 1_H \otimes \nabla_B.
	\end{equation}
	Explicitly,
	 \begin{equation}\label{FedosovConnectionComplete}
		D = d - \frac i\hbar \big(\theta + \omega_{\alpha\beta}\hat y^\alpha dy^\beta\big)
			- \frac i{2\hbar}\Gamma_{\alpha\beta\gamma} \hat y^\alpha \hat y^\beta dy ^\gamma 
			- \frac i{8\hbar}R_{\alpha\beta\gamma\delta} \hat y^{\{\alpha} \hat y^\beta \hat y^\gamma dy^{\delta\}}  + \mathcal O(\hbar ^{1})
	\end{equation}
	Now it makes sense to consider the equation $D\phi = 0$ for sections $\phi \in \Gamma(\mathfrak H\otimes B)$. The connection $D$ also induces a covariant derivative on the sections of the bundle of linear operators on $\mathfrak H$. For an observable $f \in C^\infty(M)$ we define a quantum operator $\tilde f \in  \Gamma(L(\mathfrak H))$ by the constraints 
	\begin{equation}
		D\tilde f = 0, \quad \text{and} \quad [\tilde f_y,\ \tilde g_y] = \{f,g\}(y) \mathbb 1_{\mathfrak H_y} + \mathcal O(\hbar^{1/2}) \quad \forall y \in M,
	\end{equation}
	where $\{f,g\} := \omega^{\alpha\beta} \partial_\alpha f \partial_\beta g$ is the Poisson bracket. Fedosov's recursive formula for $D^{\mathfrak H}$ allows us to also determine the form of $\tilde f$ recursively:
	\begin{equation}
		\tilde f = f + \partial_\alpha f \hat y^\alpha + \frac 12\big(\partial_\alpha\partial_\beta - \Gamma^\gamma _{\alpha\beta} \partial_\gamma \big)f \hat y^\alpha \hat y^\beta + \mathcal O(\hbar^{3/2}).
	\end{equation}	
	This is Fedosov's generalization of the Weyl quantization rule \eqref{OperatorTrafoAlpha}. Note that the expression $(\partial_\alpha\partial_\beta - \Gamma^\gamma _{\alpha\beta} \partial_\gamma \big)f \hat y^\alpha \hat y^\beta$ is the image of $\nabla df \in \Gamma(Sym^2(T^*M))$ under the Weyl representation $dy^\alpha \to\ \hat y^\alpha$, hence is independent of the selected coordinates. Choose an arbitray point $y \in M$, then the Schrödinger equation for a wave function $\psi_y \in \mathfrak H_y \otimes B_y$ reads
	\begin{equation}
		i\hbar \partial_t \psi_y(t) = \tilde H_y \psi_y(t).
	\end{equation}
	We can extend the wave function $\psi_y$ from the single fibre $\mathfrak H_y$ to a parallel section $\psi \in \Gamma(\mathfrak H \otimes B)$ by defining $\psi_z = U(y,z) \psi_y$, where $z \in M$ and $U(y,z)$ is the parallel transport operator associated to the connection $D$. Let $c: I\to M$ be a solution to the Hamilton equation, like in Appendix A we can consider the time-dependent parallel transport $\phi(t) = U(y,c(t)) \psi_y(t)$. The Schrödinger equation formulated in terms of $\phi$ is:
	\begin{equation}
		i\hbar \partial_t \phi = \big(\tilde H_{c(t)} - i\hbar A_{c(t)} (\dot c(t)) \big) \phi(t), 
	\end{equation}
	where $A$ is the connection form of our connection $D$, see \eqref{FedosovConnectionComplete}. Explicitly evaluating $A$ on $\dot c$ gives
	\begin{equation}
	  -i\hbar A(\dot c) = -\theta_\alpha\omega^{\alpha\beta}\partial_\beta H - \partial_\alpha H \hat y^\alpha 
		+\frac 12 \Gamma^\gamma_{\alpha\beta}\partial_\gamma H \hat y^\alpha \hat y^\beta 
		- \frac 18  R_{\alpha\beta \gamma\delta}\omega^{\delta\kappa}\partial_\kappa H \hat y^\alpha \hat y^\beta \hat y^\gamma + \mathcal O(\hbar^2).
	\end{equation}
	and inserting this into the Schrödinger equation:
	 \begin{equation}\label{SchroedingerCurvedFinal}
		  i\hbar\partial_t \phi  = \Big(H-\theta_\alpha\omega^{\alpha\beta}\partial_\beta H +\frac 12  \partial_\alpha\partial_\beta H \hat y^\alpha \hat y^\beta +\mathcal O(\hbar^{3/2})\Big) \phi.
	 \end{equation}
	Up to this order in $\hbar$, the equation looks exactly like in the flat case, but this will not be true for higher orders. Since the right hand side of \eqref{SchroedingerCurvedFinal}
	does not contain any terms linear in the $\hat y^\alpha$, we can deduce again that the $\hbar$-filtration is preserved under the time evolution, and for the expectation value of $\tilde f$ we get
	\begin{equation}
		\langle \tilde f \rangle_\phi (t) = f(c(t)) + \underbrace{\partial_\alpha f(c(t)) \langle \hat y^\alpha\rangle_{\phi(t)}}_{\mathcal O(\hbar ^{1/2})} + \frac 12\underbrace{ \big(\partial_\alpha\partial_\beta - \Gamma^\gamma _{\alpha\beta} \partial_\gamma \big)f \langle \hat y^\alpha \hat y^\beta \rangle_{\phi(t)}}_{\mathcal O(\hbar ^1)} + \mathcal O(\hbar^{3/2}),
	\end{equation}	
	which in the classical limit converges to its classical value: $\lim_{\hbar \to 0} \langle \tilde f \rangle_\phi (t) = f(c(t))$.

\bigskip

	This approach of representing quantum mechanical states by means of parallel sections of a metaplectic spinor bundle was proposed in \cite{Reuter98}, although the pre-quantum line bundle is still missing from the construction there (except in the flat case). The results of the present paper have been presented first in \cite{Noelle10}, where they are derived in a top-down approach starting from the geometrical construction of this section. This paper is an attempt to present the results in a bottom-up approach instead, in order to make them more accessible.

\section*{Acknowledgements}

	I am grateful to Sebastian Krug for reading the draft of this document and pointing out an inconsistency regarding the interpretation of the results in the classical limit.

\end{document}